# Highly-chirped Bragg gratings for integrated silica spectrometers

**JAMES W. FIELD,**[*] **SAM A. BERRY,** **REX H. S. BANNERMAN,** **DEVIN H. SMITH, CORIN B. E. GAWITH, PETER G. R. SMITH,** AND **JAMES C. GATES**

*Optoelectronics Research Centre, University of Southampton, University Road, Southampton SO16 1BJ, UK*
[*]*j.field@soton.ac.uk*

**Abstract:** A blazed chirped Bragg grating in a planar silica waveguide device was used to create an integrated diffractive element for a spectrometer. The grating diffracts light from a waveguide and creates a wavelength dependent focus in a manner similar to a bulk diffraction grating spectrometer. An external imaging system is used to analyse the light, later device iterations plan to integrate detectors to make a fully integrated spectrometer. Devices were fabricated with grating period chirp rates in excess of 100 nm mm$^{-1}$, achieving a focal length of 5.5 mm. Correction of coma aberrations resulted in a device with a footprint of 20 mm×10 mm, a peak FWHM resolution of 1.8 nm, a typical FWHM resolution of 2.6 nm and operating with a 160 nm bandwidth centered at 1550 nm.

## 1. Introduction

The miniaturisation of spectrometers has been an important topic for over 20 years [1–10]. In general these spectrometers fall into two categories: Fourier transform and dispersive. A Fourier transform spectrometer utilises interference patterns, wherein the Fourier transform of the recorded interferogram gives the power spectrum. In contrast, a dispersive spectrometer uses an optically dispersive element to spatially separate the input spectrum. The input spectrum can be measured from the recorded light distribution across the sensor. It is this dispersive regime that will be investigated in this paper.

Optical sensing is an attractive application for integrated spectrometers; Bragg gratings in optical fiber and planar platforms can be used to monitor a variety of parameters including temperature, strain, chemical concentrations and pressure [3,11,12]. These sensors change their spectral response (often the position or width of a spectral peak) due to a change in external conditions. If this spectral peak is spread over many sampling bins then statistical analysis can detect changes at a greater precision than the sampling resolution. For example, the centre wavelength of a reflection peak from 1 nm bandwidth Bragg grating sensors can be tracked to below 1 pm precision using a spectrometer with a resolution of only 0.1 nm [11]. Integrated spectrometers can satisfy these resolution requirements while having a bandwidth of >100 nm [6], allowing multiplexed interrogation of tens of sensors in a single miniaturised device.

Many schemes have been used to make integrated dispersive spectrometers: arrayed waveguide gratings [4], echelle gratings [5], planar holograms [6], complex interference processes [7,8] and blazed chirped Bragg gratings [9,10,13,14]. Aside from the blazed Bragg grating schemes and complex interference schemes (which have inherent crosstalk and noise), all of these devices use etching to define phase sensitive devices; this exposes them to crosstalk due to the tight fabrication tolerances on uniformity. In contrast, Bragg grating approaches can be fabricated using UV inscription, where the refractive index can be controlled precisely and accurately.

This allows fabrication of complex devices with minimal scatter, enabling high sensitivity and dynamic range (which is important for single photon spectroscopy or Raman spectroscopy).

In this paper we fabricate and model spectrometers using blazed chirped Bragg gratings in an integrated platform. Blazed Bragg gratings have grating planes at an angle to the waveguide propagation axis which eject light from the waveguide into a plane wave propagating at an angle to the waveguide (see Fig. 1). The angle at which the plane wave travels is dependent on the wavelength of the input light, similar to a bulk diffraction grating. Chirping the grating period adds curvature to the wavefront, focusing the light in a manner similar to curved diffraction gratings. Previous devices [9,10] used an etched waveguide with a UV-induced chirped grating with a small chirp (8.75 nm mm$^{-1}$). This resulted in a device with a long focal length (>100 mm). The grating diffracted light horizontally out of the etched waveguide through an air cladding layer into a planar slab waveguide. The light focused and propagated in the horizontal plane due to vertical confinement from the planar slab.

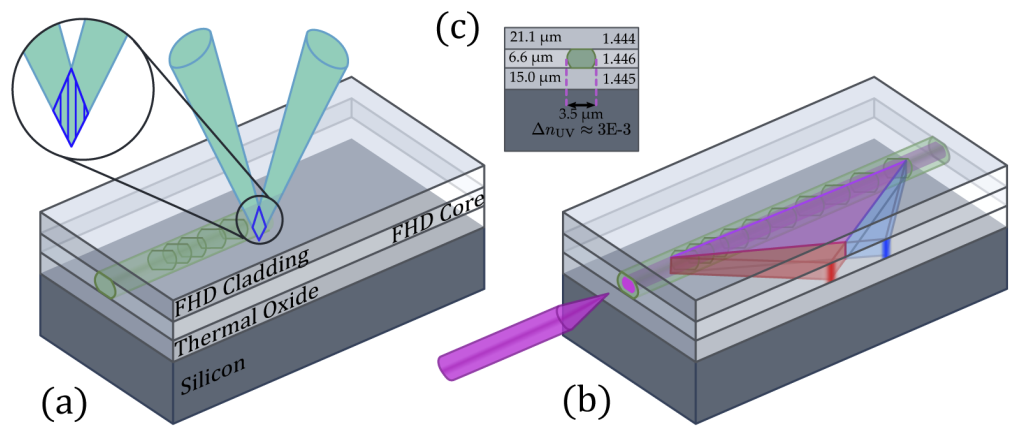

**Fig. 1.** (a) Silica-on-silicon substrate, showing UV channel waveguide and grating inscription inside a planar slab waveguide. Inset zoom shows the interference pattern used to create gratings via small-spot direct UV writing. (b) A blazed chirped grating diffracts input light (pink), into its spectral components (red and blue), while focusing the component wavelengths to different positions along the facet. The red and blue beams are vertically guided in the planar slab mode while focusing occurs. (c) Refractive index and thicknesses of layers. UV writing induces index change in the core layer of ≈3×10$^{-3}$ to form the waveguide. The index change has a Gaussian horizontal profile with a 1/e diameter of 3.5 µm.

We propose a scheme using the small-spot direct UV laser writing method [15] to embed a waveguide and 45° blazed chirped Bragg grating inside a planar slab waveguide (Fig. 1). The 45° grating only interacts with one polarisation mode (due to Brewster's angle), this largely eliminates birefringence-based crosstalk seen in [9]. Small-spot laser writing can fabricate gratings with period chirp rates in excess of 200 nm mm$^{-1}$, which allows focal lengths below 3 mm. Such devices are typically fabricated using 10 mm×20 mm chips, with gratings lengths of approximately 1 mm. The multi-layer silica-on-silicon substrate allows simultaneous creation of channel waveguides and Bragg gratings inside the planar slab waveguide inherent to the substrate. As such the loss between the channel and planar waveguide is minimized. The laser writing process fabricates devices in minutes using software control; neither phase masks nor cleanroom processing are required. This reduces the cost per device and the fabrication time. In this paper we derive some approximate figures of merit for these devices and use the beam propagation method to compare these figures of merit to full numerical simulations. Aberrations shown in the numerical model are discussed and resolved, and a device is fabricated using the results of numerical modeling.

## 2. Theory

### 2.1. Grating focal length

A blazed Bragg grating is the waveguide equivalent of a blazed reflection grating; light incident on the blazed Bragg grating is reflected at an angle dependent on both the period of the grating and its wavelength (Fig. 2(a)). Note that we refer to period, $\delta$, as the distance between grating planes as measured along the waveguide axis following the normal convention for blazed diffraction gratings. The diffraction angle is such that the phase delay between wavelets reflected from consecutive grating planes is an integer multiple of $2\pi$, which results in the following relationship with period:

$$m\lambda = n_e \delta(z) \left[1 + \cos \theta_o(z)\right], \quad (1)$$

where $m$ is the diffraction order, $\lambda$ is the wavelength of the input light, $n_e$ is the effective index of the waveguide, $\theta_o$ is the angle between the reflected light and the incident light, and $\delta(z)$ is the grating period as measured along the waveguide axis at a distance $z$ along the waveguide (see Fig. 2(a)). The grating period is related to the perpendicular distance between grating planes, $\Lambda$, and the blaze angle, $\theta_B$, by:

$$\delta(z) = \frac{\Lambda(z)}{\cos \theta_B(z)}. \quad (2)$$

In all further discussions only the case of $m = 1$ will be considered. Changing the grating period across the grating adds curvature to the wavefront reflected from the waveguide and can be used to create a focus. The focal length can be calculated using ray constructions (Fig. 2(b)). Provided that the grating length changes slowly $d\delta/dz \ll 1$, we can consider small sections of grating with approximately constant period, allowing us to calculate an effective focal length from the intersection of rays diffracted from grating sections. If the focal length is long, then $f, g \gg z$ and $f \approx g$. In this limit $\alpha$, the angle between rays from adjacent regions, is a small angle and the focal length is given by $f$:

$$\frac{f}{\sin \theta_2} = \frac{z_2 - z_1}{\sin \alpha} \approx \frac{z}{\alpha}, \quad (3)$$

$$f = \frac{z \sin \theta_2}{\alpha}. \quad (4)$$

$\alpha$ can be calculated using Eq. (1):

$$\alpha = \theta_1 - \theta_2 \quad (5)$$

$$\cos \theta_2 - \cos \theta_1 = \frac{\lambda}{n_e} \left(\frac{1}{\delta_1 - \Delta \delta} - \frac{1}{\delta_1}\right), \quad (6)$$

$$\cos \theta_2 - \cos \theta_1 = \cos(\theta_1 - \alpha) - \cos \theta_1 \approx \alpha \sin \theta_1, \quad (7)$$

where the small-angle approximation has been used to simplify Eq. (7), and $\Delta \delta = \delta_1 - \delta_2$. Provided that $\Delta \delta \ll \delta$ everywhere, Eq. (6) can be approximated:

$$\alpha \approx \frac{\lambda \Delta \delta}{n_e \delta_1^2 \sin \theta_1}, \quad (8)$$

$$f \approx \frac{n_e \Lambda^2 \sin^2 \theta_o}{\lambda \cos \theta_B C}, \quad (9)$$

where the earlier small angle approximation of $\alpha$ has been used to give $\sin \theta_1 \approx \sin \theta_2 \approx \sin \theta_o$, and $C$ is defined as $d\Lambda/dz$. Note that there is a difference in the power of $\sin \theta_o$ between this equation and that given in [10], we believe this to be an error in the original publication. If we assume a fixed $z_1$ at the start of the grating and use all the approximations stated above, the change in period with distance for a fixed focal length is linear, coinciding with previous work [9].

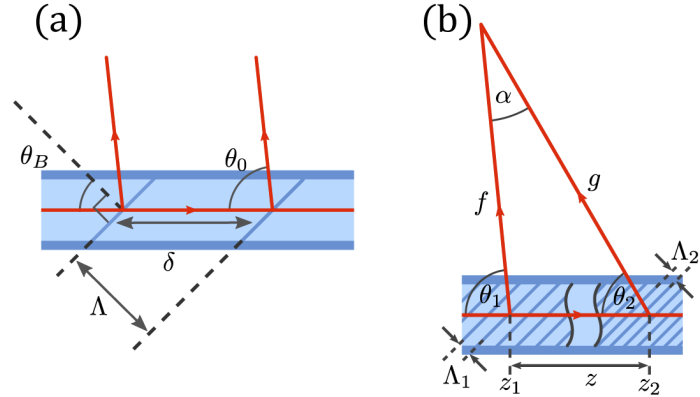

**Fig. 2.** (a) Schematic of a blazed grating showing naming conventions for grating parameters. Input light is ejected into a plane wave propagating at an angle $\theta_o$ by a grating. The solid red lines represent the direction of scattered waves, and as we will refer to later, can be described as rays. (b) Rays diffracted from sub-sections of a chirped grating forming a focus.

### 2.2. Device resolution

Much discussion exists about suitable definitions of resolution [16]. In this work resolution shall be defined as the ratio of full-width half maximum (FWHM) of the central lobe to the spot displacement with a change in wavelength (the dispersion) as measured in the detector plane. For a blazed grating spectrometer the resolution can be calculated using the focal spot diameter, $2w_f$, the focal length, $f$, and the angular dispersion with wavelength, $d\theta_o/d\lambda$. The change in $\theta_o$ along the grating is assumed to be small, which is satisfied provided the focal length is much longer than the grating length.

The diameter of the focal spot of a Gaussian apodized grating can be approximated by twice the focal waist, $w_f$, of a Gaussian beam, with the initial beam diameter given by the length of the grating, $L_g$, projected along the direction of the focus. Use of a Gaussian beam includes diffractive effects in the resolution by changing the focal spot size with the focal length. The spot size of such a beam is described by:

$$2w_f = \frac{4\lambda f}{\pi L_g n_e \sin \theta_o}. \tag{10}$$

A full derivation of the focal waist of a Gaussian beam can be found in [17]. The linear dispersion of the device (the distance the focal spot moves with a change in wavelength) can be calculated from the angular dispersion and the focal length:

$$\frac{dz}{d\lambda} = f \frac{d\theta_o}{d\lambda} = \frac{f}{n_e \delta \sin \theta_o}. \tag{11}$$

The resolution $\Delta\lambda_{\text{FWHM}}$ can be obtained using Eq. (10) and Eq. (11):

$$\Delta\lambda_{\text{FWHM}} = 2w_f \sqrt{\frac{1}{2}\ln 2} \frac{d\lambda}{dz} = \frac{2\sqrt{2\ln 2}\lambda^2}{n_e \pi L_g (1 + \cos\theta_o)}, \tag{12}$$

Where the $\sqrt{\ln 2}$ factor has been used to convert from $1/e$ to FWHM resolution. The resolution of the device is thus only dependent on $L_g$, $\theta_o$ and $\lambda$. This leaves the focal length as a free parameter which can be adjusted to ensure the dispersion of the device matches the pixel pitch on the detector array without compromising the resolution.

### 2.3. Beam propagation modeling and aberration correction

Equation (9) makes a number of assumptions about focal length and chirp rate. If a short focal length device is to be made, it is necessary to have an understanding of how the behaviour of the device differs from the ideal when these assumptions are taken to their limits. It is also desirable to have a numerical model to optimize device operation with fewer fabrication iterations. We investigate this using the beam propagation method. The beam propagation method allows the electric field at any plane to be calculated given the electric field at a starting plane. In the case of a blazed chirped grating we can propagate the near-field of the grating in a direction perpendicular to the waveguide to investigate the focus.

In 1D we can consider the Bragg grating as a sum of weak point scatterers. The amplitude of the scattered electric field from such a scatterer is proportional to its refractive index. As all scatterers are weak the amplitude of the total scattered electric field is the sum of the scattered field from each individual scatterer. The electric field close to the grating (the initial field for the beam propagation method) is then proportional to the grating refractive index contrast. In addition to the amplitude contributed by the scatterers, the phase of this near-field must also be considered. As the input field to the grating is a channel waveguide mode, it will accumulate a phase proportional to the distance traveled along the waveguide. The resulting initial field for the beam propagation method is given by:

$$E(z, x = 0) = A(z) \sin[\phi(z)] \exp(-ik_0 n_e z), \tag{13}$$

where

$$\phi(z) = \int_0^z \frac{2\pi}{\delta(z')} dz', \tag{14}$$

$x$ is the propagation distance perpendicular to the waveguide, $A(z)$ is the grating strength envelope along the waveguide (hereafter referred to as the 'apodization profile') and $k_0$ is the vacuum wavenumber. $\phi$ is the propagation phase accumulated inside the waveguide. This approach does not account for the direction of the grating planes. The distance between scatterers provides a phasematching condition for grating output direction, however, considering finite width grating planes gives a k-vector condition for both the direction and diffraction efficiency. As Fig. 3 shows, this should be a distribution about the specular reflection condition, $\theta_o = 2\theta_B$, however, the width of this distribution is unknown. In the scope of this paper we will ignore the directional effects of grating planes due to blaze angle (for a fixed wavelength) leaving more in-depth analyses for future investigation.

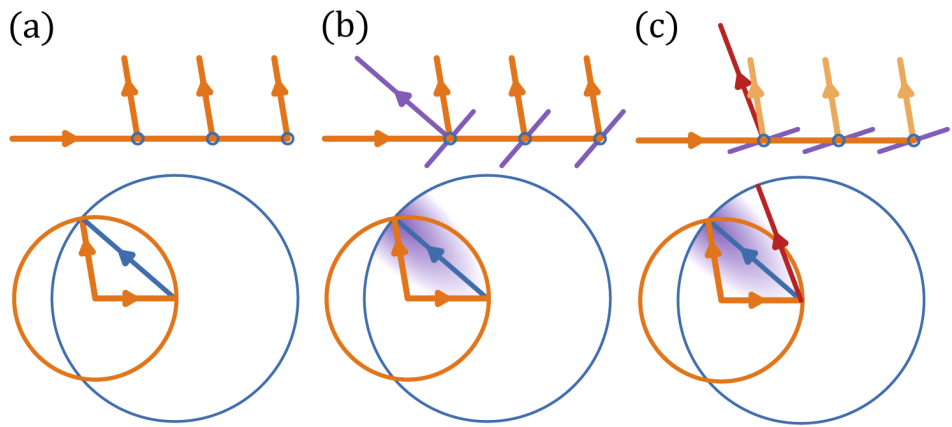

**Fig. 3.** The direction and strength of outcoupling from gratings is set by a k-vector matching condition. (a) Point scatterers diffract light in a direction dictated by phasematching (see Eq. (1)). (b) For finite width grating planes the direction normal to the grating plane should bisect the input and output rays for maximum diffraction (specular reflection). (c) If the grating blaze angle does not satisfy specular reflection at a given wavelength and $\theta_o$, it will diffract at lower efficiency.

Using the beam propagation method, the intensity distribution through the focus can be determined. A linear period chirp with distance gives a focus (Eq. (9)); for focal lengths much longer than the grating length this works as expected, but for shorter focal lengths aberrations can be seen (Fig. 4(a-c)). These will make the spectrometer deviate from the ideal resolution in Eq. (12), reducing the resolving power as aberrations become more prevalent.

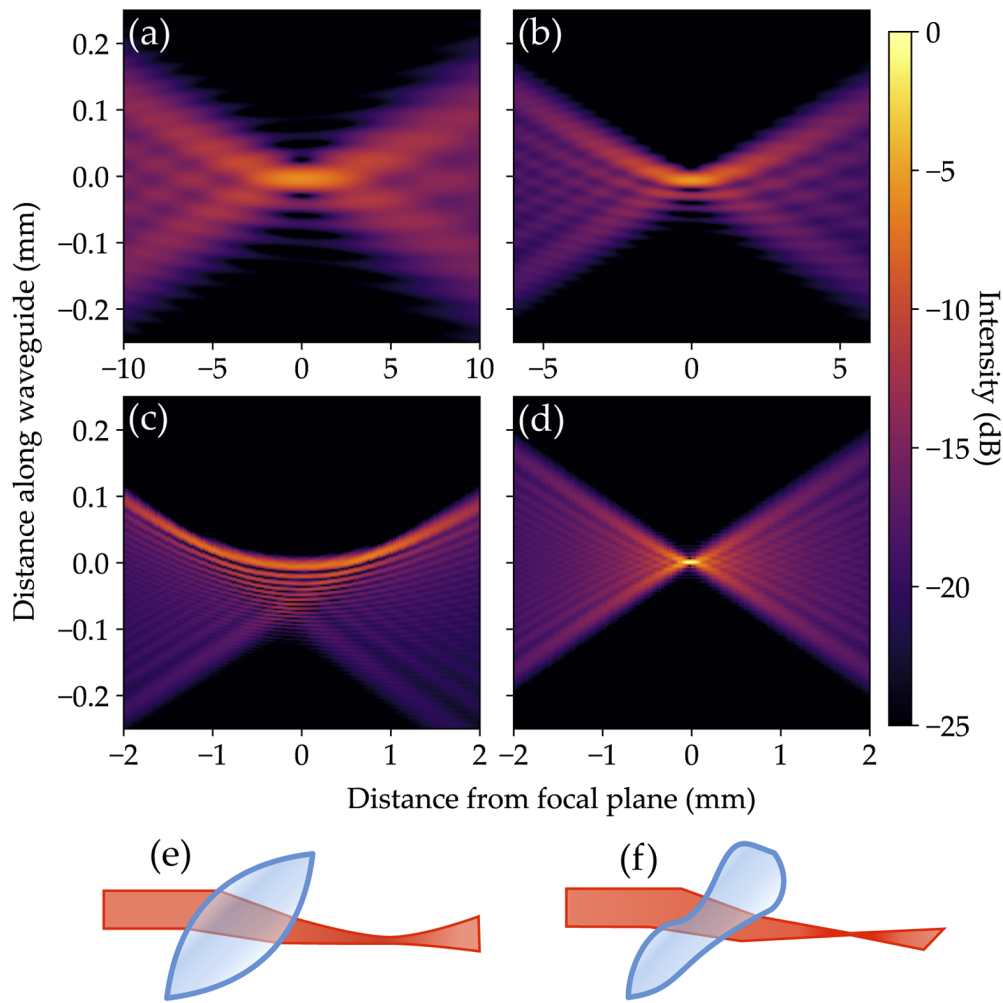

**Fig. 4.** Beam propagation modeling of 2 mm uniform apodized gratings with different periods. Horizontal scale modified with focal length to keep diffraction visually constant. (a) 50 mm focal length linear chirp. (b) 30 mm focal length linear chirp. (c) 10 mm focal length linear chirp (d) 10 mm focal length ideal chirp. (e & f) Illustrative equivalent lens system for linear (uncorrected) chirp and coma corrected chirp respectively. A perfect lens results in coma, whereas a nonlinear chirp function results in a perfect focus.

A perfect non-aberrated focus has a quadratic phase front. The phase front across the grating is given by Eq. (14). For a linear period chirp $\delta(z) = \delta_0 + Cz/\cos\theta_B$, where $\delta_0$ is the average period, Eq. (14) becomes

$$\phi(z) = \frac{2\pi \cos\theta_B}{C} \ln\left(1 + \frac{C}{\delta_0 \cos\theta_B} z\right), \quad (15)$$

which, in the limit of long focal length compared to grating length ($C/(\delta_0 \cos\theta_B) \ll L_g$) simplifies to

$$\phi(z) = \frac{2\pi}{\delta_0} z - \frac{\pi C}{\delta_0^2 \cos\theta_B} z^2. \quad (16)$$

Eq. (9) can be used to replace $C$, resulting in:

$$\phi(z) = \frac{2\pi}{\delta_0} z + \frac{k_0 n_e z^2}{2f} \sin^2\theta_o. \quad (17)$$

While Eq. (17) has a quadratic focusing term it also has a linear phase term. A linear phase across a beam represents a beam traveling at an angle to the propagation axis; a beam passing through a lens at an angle will exhibit coma aberration. This is visible as a one-sided set of fringes as can be seen in Fig. 4(c). This aberration can be corrected (see Fig. 4(d)) using a nonlinear chirp that constrains all rays to pass through a single point. This was previously suggested by Madsen et al. [9], however they were unable to fabricate or model the output of such a grating. Such a period is given by:

$$\delta(z) = \frac{\lambda_0 \cos\theta_B(z)}{n\left[1 + \frac{z - f_z}{\sqrt{(z-f_z)^2 + f_x^2}}\right]}, \quad (18)$$

where $f_x$ and $f_z$ are the components of the vector from the centre of the grating to the focal point projected along the waveguide axis and perpendicular to the waveguide axis, respectively. Moving away from the design wavelength results in coma, with the degree of coma increasing as the wavelength gets further from the design wavelength. Such a complex nonlinear chirp can be fabricated using the small-spot direct UV writing (DUW) method described in the next section.

## 3. Method

Devices were fabricated to test coma correction and resolution predicted from modeling. A silicon wafer with a 15 µm thermal oxide layer was used as the substrate for the device. Doped glass layers are deposited on top of the thermal oxide using flame hydrolysis deposition (FHD); the first of these is a germanium doped photosensitive core, the second is a non-photosensitive cladding that matches the index of the thermal oxide layer. The refractive index, photosensitivity and thickness of the core layer (parameters of which can be found in Fig. 1(c)) are engineered to ensure there is a strong single-mode planar slab waveguide, and a single-mode channel waveguide (after UV exposure) with a typical effective refractive index of 1.448.

Channel waveguides and Bragg gratings are fabricated using the DUW method [18]. Two beams from a 244 nm frequency-doubled argon-ion laser are interfered at their mutual focus to create an interference pattern, with the period of the pattern set by the angle between the beams and the wavelength of the writing laser.

Phase control of one of the beams via an electro-optic modulator (EOM) shifts the interference fringes with respect to the focal spot; this forms Bragg gratings by holding the fringes stationary with respect to the waveguide while the spot moves along the chip [15]. Gratings of a different period can be written by mismatching the shift rate and the writing speed; this method can also be used to fabricate chirped gratings. Gratings written with a matched shift rate have the highest possible strength; refractive index modulation depth of gratings decreases as gratings are written further from the interferometer design period. The range over which the system can write gratings in a single device depends on the design period of the interferometer and the size of the writing spot. The spectral dependence on achievable grating strength is a Gaussian in frequency offset from the frequency of the interference period, which is defined by the angle of beam interference (for a full discussion, see Ref. [19]). This is centered at a period of ≈760 nm with a FWHM of ≈76 nm for the system used to fabricate the device. The relation between achievable grating strength and detuning from the design period of the interference pattern is referred to as the 'detuning profile' [15]. Parameters of desired gratings are defined in software which interfaces with an Aerotech A3200 nanometer precision stage system to produce voltages for the phase control system, allowing near arbitrary phase control of gratings. A full discussion of this fabrication platform and fabrication tolerances can be found in [19]. A 1 mm grating, with a blaze angle of 45°, a uniform apodization profile and a nonlinear period chirp was fabricated in a direct UV written waveguide in FHD silica. The effective index of the waveguide has been measured in non-blazed samples to be approximately 1.4481. The grating period was designed to focus light at a wavelength of 1550 nm (without aberrations) 5.55 mm from the waveguide. The waveguide was written 0.65 mm from the edge of the chip, focusing light in free space outside the chip; Eq. (18) was modified to account for refraction at the chip facet.

Devices were characterized using an Agilent 81600B tuneable laser and a Raptor Photonics OW1.7-VS-CL-640 camera. The linewidth of the tuneable laser was much narrower than the resolution of the spectrometer, so could be used to probe the response of the device over the operating bandwidth. The laser was coupled into the chip using a single mode 1550 nm polarisation-maintaining fiber v-groove (Fig. 5). The chip facet was not polished, so a 100 µm thick cover slip was used to provide a smooth facet on the side of the chip, with refractive index matching oil between the chip and the cover slip. The camera was aligned with the axis of the imaging system perpendicular to the waveguide and images were taken through a 57× magnification system. This decreased the effective pixel pitch of the image, allowing the focus formed by the grating to be probed at a higher spatial resolution than the camera would normally allow. The distance over which the intensity pattern moved was much larger than the size of the detector (due to the magnification system); the camera was translated, and its position was recorded using a digital micrometer attached to the translation stage. Repeat wavelength measurements were taken every time the camera was moved to ensure the measured offset was correct; aberrations from being near the edge of the imaging system were observed to be minimal. The focus occurred outside the waveguide, so whilst the spectrometer output had a narrow, focused output in the horizontal axis it had diffracted in the vertical axis (see Fig. 6). This vertical stripe was larger than the front lens in the imaging system, so created a lens flare in the image; visible as a line in the centre of all images with a small angle to the vertical axis. Data was collected such that the lens flare never obscured the spectrometer peak. Fig. 6 shows a series of unprocessed images obtained by tuning the laser source to different wavelengths. Columns of grayscale pixel intensity values were summed to reduce the data to a 1D array. The pixel number was converted to a physical position using the pixel pitch, magnification and camera position, such that the position dependent intensity distribution for different wavelengths and camera positions could be characterised. Away from the design wavelength, sidelobes started to interfere with the fit of the central peak, so a basic fitting routine was designed to fit to the central peak irrespective of sidelobes. Central peak and sidelobe positions were found using a peak prominence algorithm. A Gaussian was fitted to each peak, optimizing the least square deviation of the data from the sum of all the Gaussians together. The centre position and width of the Gaussian fit for the central peak were extracted. The intensity distribution covered over 40 pixels, fitting allowed small changes in the centre position (far less than the peak width) to be measured. This fitting routine successfully detected each peak, though the extra degrees of freedom meant that far from the design wavelength (where the signal to noise ratio was lower) the variability in the fitting of the central peak increased. The same fitting routine was used to extract data from modeled devices.

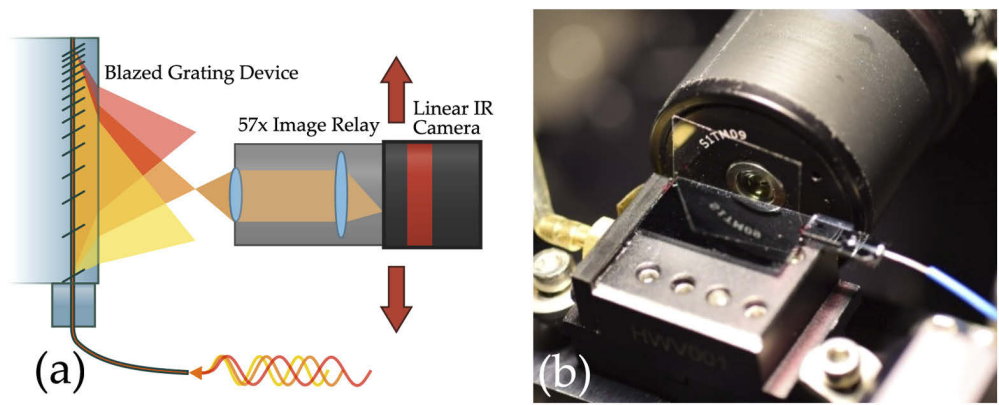

**Fig. 5.** (a) Camera system used to image spectrometer output as wavelengths were changed. A magnification system was used to better resolve the spatial intensity profile of the focus. A micrometer stage was used to move the camera and imaging system along the edge of the chip. (b) Fabricated device and imaging system. Images were taken through a silica cover slip, as the chip facets were not polished.

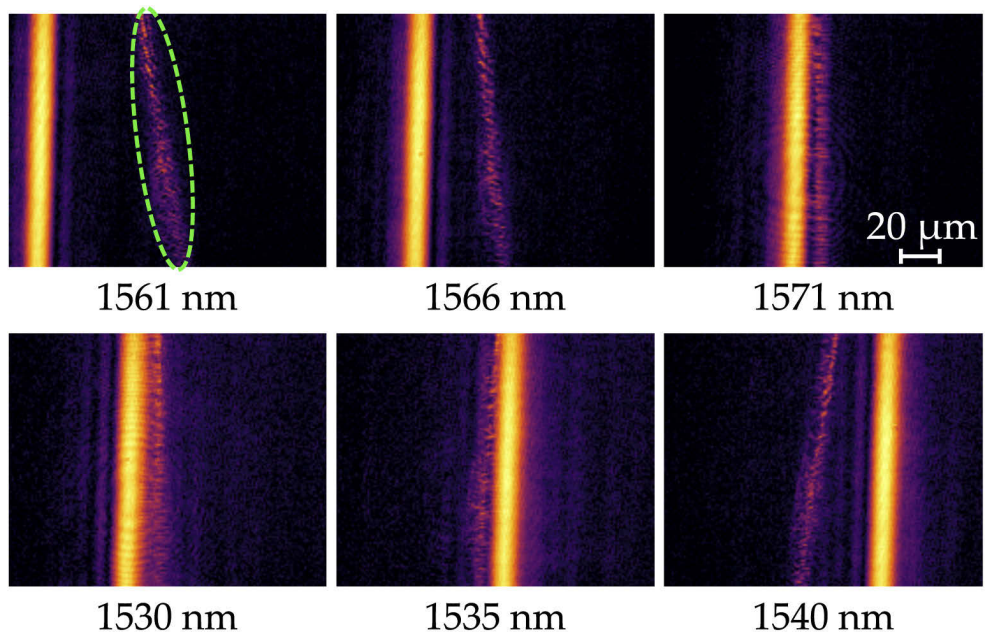

**Fig. 6.** Series of camera data of spectrometer output with intensities displayed in the log domain. Images on the same row were taken with the same camera position. Diffraction from propagation outside the chip causes the distribution vertically; the vertical distribution overfills the lens of the imaging system, resulting in lens flare (highlighted by green ellipse), which changes angle depending on the position of the maximum of the horizontal distribution. Weak sidelobes are visible in the top row on the right of the main spectrometer output and stronger sidelobes are visible on the bottom to the left of the output. The sidelobes are on different sides as the top and bottom row are respectively above and below the design wavelength. At the design wavelength sidelobes disappear.

## 4. Results and analysis

The focal length of the device was measured by imaging the planar slab mode and translating the imaging system away from the device until the output was focused. The focal length was estimated to be between 5.4 mm and 5.7 mm from the waveguide. Images of the output intensity profile were taken with the imaging system focal plane 5.50 mm from the waveguide (see Fig. 6). Images showed a vertical stripe which moved as the input wavelength was changed. Sidelobes developed in the camera images as the input wavelength was tuned away from the design wavelength (see Fig. 8). Beyond the edges of the wavelength range used to collect data the signal-to-noise ratio of the device decreased and fitting became highly variable. A clear profile could still be observed; a more robust fitting algorithm could allow device characterisation over a wider bandwidth. The intensity pattern at the focal plane was interrogated and the position and width of the central lobe were extracted using the described fitting routine. This was compared to a model of the device using the beam propagation method and the parameters of the fabricated grating. The intensity profile was extracted 5.54 mm from the modeled grating, this distance was chosen as it provided the best overlap between theoretical and modeled data. Fig. 7(a) shows experimental data and theoretical data were in close agreement about the position of the central lobe, however Fig. 7(b) shows that there was a disagreement about the width, and hence the resolution. The grating was fabricated with a uniform apodization profile; the image in the focal plane was expected to have a $sinc^2$ profile. Fig. 7(c) shows that the modeled device has a central peak and sidelobes but the fabricated device has a wider central peak and no sidelobes. This implies that the grating strength profile $A(z)$ is not uniform and decreases smoothly towards the edge of the grating (hence the lack of sidelobes).

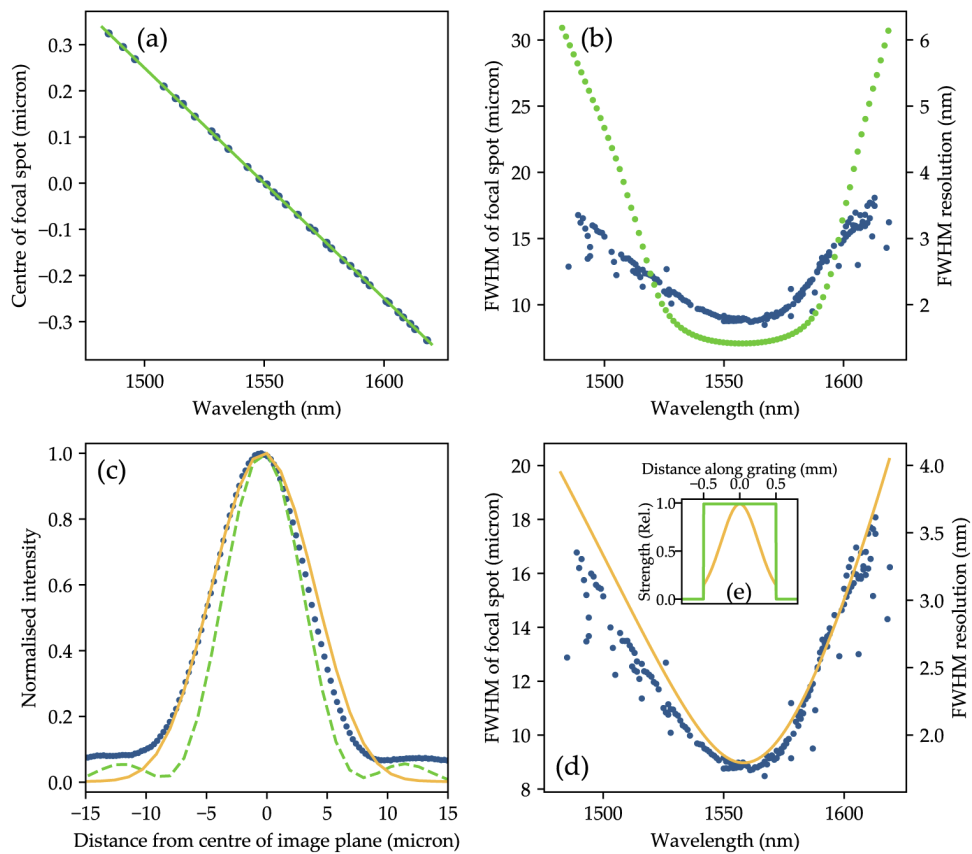

**Fig. 7.** Performance of fabricated device (blue dots), and modeled device before and after $A(z)$ modification (green and orange respectively). (a) Position of central lobe for modeled and fabricated devices. (b) Focal spot diameter and associated resolution showing disparity between modeled device (before $A(z)$ modification) and fabricated devices. (c) Intensity profile at the focal plane (at 1550 nm) showing the difference in width and sidelobes between fabrication and modeling (before and after $A(z)$ modification) (d) Resolution of fabricated device and modeling (after $A(z)$ modification). (e). Grating strength profile before and after $A(z)$ modification.

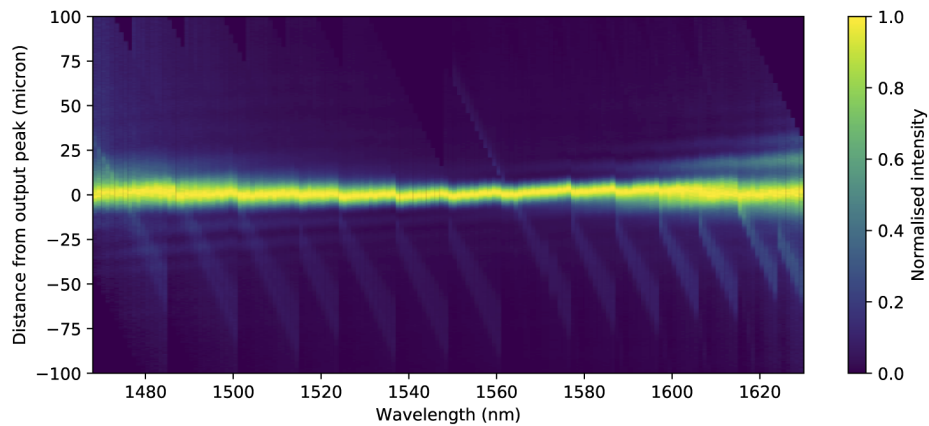

**Fig. 8.** Normalised intensity profile of device output, referenced from the centre of the intensity distribution. Lens flare can be seen as a series of diagonal lines with a negative gradient. Discontinuities are due to stitching errors, likely caused by field curvature in the imaging system. Spectrometer aberrations can be seen as an asymmetric widening of the central lobe, and sidelobes. On either side of the design wavelength aberrations appear on opposite sides of the central peak.

There are 2 known effects that could cause a reduction in the grating strength profile near the edges: the finite width of the grating detuning profile combined with the large grating chirp, and the reduction in outcoupling efficiency due to directional effects of grating planes due to blaze angle Fig. 3. The effect of the detuning profile can be included in the $A(z)$ term, however the effect on diffraction efficiency due to blaze angle is not well understood. The model was changed to include the functional form of the detuning profile in $A(z)$ and the width of the detuning profile was varied to achieve the best overlap with the resolution curve of fabricated devices (see Fig. 7(d)). The best overlap was achieved with a detuning width of 62 nm (see Fig. 7(e)), however the detuning width of the system used to fabricate the gratings is ≈76 nm. This implies that both the detuning profile and the directional effects of blaze angle were reducing the strength of the edge of the grating.

After modifying $A(z)$ the resolution of the modeled and fabricated devices were in good agreement. The peak spectral resolution was 1.8 nm and over a 100 nm range the spectral resolution was below 3 nm. The peak resolution occured at 1560 nm, instead of the design wavelength of 1550 nm, however modeling showed this can result from an error in the focal plane distance. The peak resolution was in good agreement with the value predicted by Eq. (12) (1.75 nm).

## 5. Discussion

This paper lays the foundations of an integrated spectrometer scheme in an integrated silica platform with >160 nm bandwidth. The resolution is lower than that achieved previously; Madsen et al. [9] showed a FWHM resolution of 0.15 nm using a 10 mm grating, though their bandwidth was only 7.8 nm. The advantage of the scheme and fabrication method presented here is the flexibility and breadth of the parameter space that can be varied.

We have shown the ability to remove the effects of coma aberrations, allowing for small footprint devices without compromising the resolution. Reducing the focal length does worsen the effects of aberrations at wavelengths away from the design wavelength, however historically in diffraction grating spectrometers this aberration has been removed using different detector mounting geometries [20].

It has been observed in the current device that the peak resolution is limited by grating length (as it satisfies the value predicted by Eq. (12)); provided it is possible to correct for other aberrations then the resolution should be limited entirely by the grating length. To increase the grating focal length while preserving the small footprint, the total grating chirp should be increased, however this then leads to further problems as the total grating chirp approaches the limit of detuning bandwidth, a hard limit set by the current fabrication method.

There are a few possible solutions to this problem. The focal length could be increased, which reduces the rate of chirp (and therefore total chirp) required to create a longer grating, however this results in a larger footprint. A phase mask system could also be used to write arbitrarily long gratings without problems with grating detuning, however this would reduce flexibility in fabricating new devices which has been immensely useful when troubleshooting current devices. Finally the required change of $\delta$ could be achieved by varying a combination of period and blaze angle ($\Lambda$ and $\theta_B$). Whilst this preserves the flexibility of the current fabrication system and the small footprint of the short focal length devices additional investigation is required to ensure that potential increases in grating length are not offset by loss of diffraction efficiency from moving away from the momentum matching condition discussed in Fig. 3.

Future work aims to investigate wavelength dependent aberrations to try and increase the resolution away from the design wavelength, as well as using a combination of blaze angle and $\Lambda$ to increase the grating length.

## 6. Conclusions

We have demonstrated a planar spectrometer platform with 1.80 nm peak FWHM resolution and 2.6 nm typical FWHM resolution, >160 nm bandwidth, fiber compatibility and the potential for high-throughput and high-resolution operation over a large bandwidth. An efficient numerical model for dispersive Bragg grating spectrometers was shown using the beam propagation method, this was used to examine and correct for aberrations due to short focal lengths. Further work aims to increase the resolution by increasing the grating length and varying the blaze angle along the grating, as well as investigate the effects of blaze angle and wavelength on grating diffraction efficiency.

**Funding**

Engineering and Physical Sciences Research Council (EP/M024539/1, EP/N509747/1).

## Acknowledgements

All data supporting this study are openly available from the University of Southampton repository at https://DOI.org/10.5258/SOTON/D1356

## Disclosures

The authors declare no conflicts of interest.